\documentclass{iopart}
\usepackage{natbib}
\usepackage{graphicx}
\usepackage{bbm}

\begin{document}

\title{Maximum Likelihood Method for Cross Correlations with Astrophysical Sources}

\author{Ronnie Jansson and Glennys R. Farrar}

\address{Center for Cosmology and Particle Physics,
Department of Physics\\ New York University, NY, NY 10003, USA\\
}


\begin{abstract}
We generalize the Maximum Likelihood-type method used to study cross correlations between a catalog of candidate astrophysical sources and Ultrahigh Energy Cosmic Rays (UHECRs), to allow for differing source luminosities.  The new method is applicable to any sparse data set such as UHE gamma rays or astrophysical neutrinos.  Performance of the original and generalized techniques is evaluated in simulations of various scenarios.   Applying the new technique to data, we find an excess correlation of about 9 events between HiRes UHECRs and known BLLacs, with a $6\times 10^{-5}$ probability of such a correlation arising by chance. 
\end{abstract}


\section{Introduction}

Correlation studies play a fundamental role in establishing or ruling out candidate sources for rare events such as the highest energy cosmic rays and UHE gamma rays. When the number of events is low, it is necessary to use sensitive methods to be able to identify the sources and quantitatively assess the possibility of an incorrect identification. The purpose of the present work is to evaluate and improve these methods.  If the hypothetical sources are commonplace rather than rare, so that the typical separation between candidate sources is not large compared to the directional uncertainty of individual events, then an entirely different approach must be used, requiring much larger datasets.  Fortunately, the GZK horizon allows the surface density of source candidates to be reduced by increasing the energy threshold; this played an essential role in the recent Auger discovery \citep{AugerAGN:2007} of a correlation between UHECRs and galaxies in the Veron-Cetty Veron Catalog of Quasars and AGNs.   Furthermore, it is plausible that only unusual astrophysical objects can produce very rare events, in which case the candidate source catalog may naturally be sparse.  

Hints of excess correlations were reported earlier between ultrahigh energy cosmic rays (UHECRs) and BLLacs by \cite{TT:BLLacs01}, \cite{TT:BLLacs04}, \cite{HR:BLLacs} and between UHECRs and x-ray clusters by \cite{pf05}. The analyses in  \cite{TT:BLLacs01}, \cite{TT:BLLacs04}, and \cite{pf05} determine the number of correlated events within some given angular separation between UHECR and source, and calculate the ``chance probability'' of finding a correlation at the observed level by doing a large number of simulations with no true correlations.  The \citep{AugerAGN:2007} analysis in addition ``scans" to find the optimal UHECR energy threshold and maximum source redshift.

In order to incorporate the experimental resolution on an event-by-event basis, \cite{HR:BLLacs} proposed a Maximum Likelihood-type procedure and applied it to studying correlations between UHECRs and BLLacs. This procedure (denoted the HiRes procedure, below) is motivated under the unphysical assumption that every candidate BLLac source has the same apparent luminosity. Even if BLLacs were standard candles with respect to UHECR emission, the BLLacs in the catalog have a large range of distances which would imply an even larger range of apparent luminosities, so one does not want to rely on such an assumption. Furthermore the validity of the method was not demonstrated via simulations.

In the present paper we introduce a ML prescription which avoids the assumption of equal apparent source luminosity and allows the potential sources to be ranked according to the probability that they have emitted the correlated UHECRs. We test and compare both methods with simulations in a variety of situations. We find that the HiRes method gives the correct total number of correlated events even when the sources do not have equal apparent luminosities, as long as the numbers of events are sufficiently low and the candidate sources are not too dense or clustered themselves. We find that our new method performs better in these more challenging cases. In a final section, we apply the new procedure to BLLacs and x-ray clusters.

\section{Maximum Likelihood Approach for the Cross-Correlation Problem}\label{method}

\subsection{The HiRes Maximum Likelihood method}

In the HiRes Maximum Likelihood method of \cite{HR:BLLacs}, the aim is to find, among $N$ cosmic ray events, the  number of events, $n_s$, that are \emph{truly} correlated with some sources, of which the total number is $M$. Hence, there are $N-n_s$ background events whose arrival directions are given by a probability density $R(\bf{x})$, which is simply the detector exposure to the sky as a function of angular position, $\bf{x}$. For a true event with arrival direction $\bf{s}$, the \emph{observed} arrival direction is displaced from $\bf{s}$ according to a probability distribution $Q_i(\bf{x},\bf{s})$. Note that $Q_i$ is governed only by the detector resolution.  For the analysis given in \cite{HR:BLLacs}, $Q_i$ is taken to be a $2d$ symmetric Gaussian of width equal to the resolution $\sigma_i$, of the $i$th event. (Note that, throughout, the parameter $\sigma$ is related to $\sigma_{68}$, the radius containing 68\% of the cases, by $\sigma_{68} = 1.51\sigma$.) Dispersion due to random magnetic fields can be incorporated into the $Q_i$ by generalizing the variance to $\sigma_{effective}^2=\sigma_{detector}^2+\sigma_{magnetic}^2$, where $\sigma_{magnetic}$ may vary with direction and event energy. In practice the magnetic dispersion is not well known and has to be treated as an unknown or argued to be smaller than $\sigma_{detector}$. 

The probability density of observing the $i$th  event in direction $\textbf{x}_i$ is 
\begin{equation}
P_i(\textbf{x}_i) = \frac{n_s\sum_{j=1}^MQ(\textbf{x}_i,\textbf{s}_j)R(\textbf{s}_j)}{N\sum_{k=1}^{M}R(\textbf{s}_k)} + \frac{N-n_s}{N}R(\textbf{x}_i),
\end{equation}
and the \emph{likelihood} for a set of $N$ events is defined to be $\mathcal{L}(n_s) = \prod_{i=1}^NP_i(\textbf{x}_i)$,
which is maximized when $n_s$ is the true number of correlated events. Since $\mathcal{L}$ is a very small number, which depends on the number of events, it is more useful to divide $\mathcal{L}$ by the likelihood of the \emph{null hypothesis}, i.e., $n_s=0$, to form the likelihood ratio $\mathcal{R}(n_s) = \mathcal{L}(n_s)/\mathcal{L}(0)$. The logarithm of this ratio is then maximized to obtain the number of correlated events, $n_s$. The significance of the correlation is evaluated by measuring the fraction $\mathcal{F}$ of simulated isotropic data sets with as large or larger value of $\ln \mathcal{R}$.

\subsection{Extending method to differing luminosities}\label{var_lumin}

If the source candidates for UHECRs were standard candles with known distances, or if the relative fluxes from different sources were known, we could generalize the above method by simply attaching the appropriate relative weight for each source and maximizing the likelihood ratio to obtain the number of correlations, $n_s$. However, for the applications we have in mind we do not know the relative luminosities of the putative sources, nor do we expect them to be standard candles. Thus, to generalize the HiRes method to allow for sources with differing luminosities we assign an a apriori unknown number of cosmic rays, $n_j$, separately for each source, with $n_{tot} = \sum_{j=1}^Mn_j$. The probability density generalizes to
\begin{equation}
P_i(\textbf{x}_i) = \frac{\sum_{j=1}^Mn_jQ(\textbf{x}_i,\textbf{s}_j)R(\textbf{s}_j)}{N\bar{R}_s}
+  \frac{N-\sum_{j=1}^Mn_j}{N}R(\textbf{x}_i),
\end{equation}
where $\bar{R}_s=\sum_{j=1}^MR(\textbf{s}_j)/M$. As above, we divide the likelihood $\mathcal{L}$ by the likelihood of the null hypothesis, $\mathcal{L}(\{n_j\}=0)$, to obtain the likelihood ratio
\begin{equation}\label{lnR}
\mathcal{R} = \prod_{i=1}^N\left[\frac{1}{N}\left(\frac{\sum_{j=1}^Mn_jQ(\textbf{x}_i,\textbf{s}_j)R(\textbf{s}_j)}{R(\textbf{x}_i)\bar{R}_s} - \sum_{j=1}^Mn_j\right) + 1   \right].
\end{equation}
Maximizing $\ln \mathcal{R}$, with respect to the set $\{n_j\}$ of $M$ numbers, determines the most probable values of the set $\{n_j\}$ of $M$ numbers as well as the set $\{(\ln \mathcal{R})_i\}$ of $N$ numbers. The numbers $\{(\ln \mathcal{R})_i\}$ are figures of merit, providing information about how strongly correlated the individual cosmic ray events are to the catalog of sources, and allow us to rank the cosmic rays in order of their likelihood of being correlated to the source data set (similar information can be obtained in the HiRes approach from the contributions of individual cosmic rays to the sum in $\ln \mathcal{R}(n_s)$).

A crucial difference between the generalized method and the HiRes method is that in the new method $n_{tot} = \sum n_j$ provides an estimate of \emph{all} correlations, i.e., both \emph{true} and \emph{random} correlations, whereas the HiRes method yields only an estimate of the number of \emph{true} correlations. In the generalized method one  obtains $n_j>0$ if any cosmic ray event is close enough to any source, regardless of the degree of correlation between the two data sets elsewhere. For the HiRes method, the single optimized parameter, $n_s$, will be greater than zero only if the degree of correlation between the two data sets are greater than what is expected for a random sample of cosmic rays (weighted by the detector exposure). 

We now turn to the estimation of $n_s$ with the new method. A crude estimate of the number of true correlations for the new method is $n_{tot}-\bar{n}_{rand}$, where $\bar{n}_{rand}$ is the average number of correlations obtained when cosmic rays are uncorrelated to the data set of potential sources, i.e., cosmic rays drawn from an exposure-weighted isotropic distribution. 

A better measure of $n_{s}$ is summarized by the following equations, where the superscript in parentheses labels the ``order'' of refinement, giving successively better approximations to $n_{s}$:
\begin{eqnarray}
n^{(0)} &=& n_{tot} = \sum_j^Mn_j\\
n^{(1)} &=& \bar{f}^{-1}(n^{(0)} - \bar{n}_{rand}(N))\\
n^{(2)} &=& \bar{f}^{-1}(n^{(0)} - \bar{n}_{rand}(N-n^{(1)})),
\end{eqnarray}
where $\bar{n}_{rand}(N)$ is the average total number of correlations found with $N$ events drawn from an isotropic distribution weighted by the detector exposure, and where $\bar{f}$ is the average fraction of true events that are recovered as correlated. This fraction will typically be slightly smaller than unity since under the assumption that true correlations are separated according to a Gaussian distribution, some events are separated too far from their sources to be accepted as correlated events. We measure $\bar{f}$ by simulations.

Up to this point, we have not specified whether $\{n_j\}$ are real numbers or integers. Indeed,  the method allows either choice. A non-integer implementation has the virtue of encoding more information into the set of numbers, $\{n_j\}$, while in the integer case it is necessary to also study $\{(\ln R)_i\}$ in order to rank the sources in terms of correlation quality to the set of cosmic rays. An integer approach may be preferred as it provides a more concise answer as to which are the likely correlated sources. Moreover, in the case of strong correlations and ample statistics, the integer implementation is likely to offer a more intuitive result in terms of singlets, doublets, etc.. In the following sections we will use the integer implementation, unless noted otherwise. 

The procedure to maximize $\ln\mathcal{R}$ may depend on whether $\{n_j\}$ are non-integers or integers. In the present work, we use a commercial non-linear optimization package for the former case. One might expect to encounter  problems when the number parameters to be fitted ($M$) exceeds the number of data points ($N$). However, in practice this never becomes an issue as long as $M\pi\sigma_{mean}^2$ is small compared to the solid angle observed so the number of random correlations is not too large. Then, most potential sources are well separated from most UHECR events, and thus have $n_j=0$, thanks to the rapid cutoff of $Q$ for $s\gg \sigma$, where $s$ is the separation between cosmic ray and source.
For the integer case we first note that the exponential form of $Q$ and the diluteness of the source data set will make it extremely rare for a correlated event to be ascribed to the wrong source. In other words, a cosmic ray event must be at almost the exact same (small) angular distance from two potential sources for there to be any ambiguity as to from which source the event originated. For the $i$th cosmic ray, this allows us to consider only the source $j$ that maximizes the quantity $Q(\textbf{x}_i,\textbf{s}_j)R(\textbf{s}_j)$, and put that quantity to zero for all other sources. The maximization of equation \ref{lnR} is then a matter of testing, for each source, whether $\ln\mathcal{R}$ decreases when $n_j$ is increased from 0 to 1. If it does, then $n_j=0$. If $\ln\mathcal{R}$ increases, then we repeatedly increase $n_j$ by unity, until $\ln\mathcal{R}$ decreases and we find  the correct number of correlations, $n_j$, for that source.

\section{Simulation trials}\label{simulations}

In vetting the two methods with simulations, we test their ability to correctly reproduce the number of true correlations on mock data sets. This allows us to explore the effect of large event and source densities, the effect of anisotropy in the source distribution, and the consequences of having incorrect event resolutions.

\subsection{Dilute source and UHECR data sets}

\begin{figure}
 \begin{center}
 \includegraphics[scale=.75]{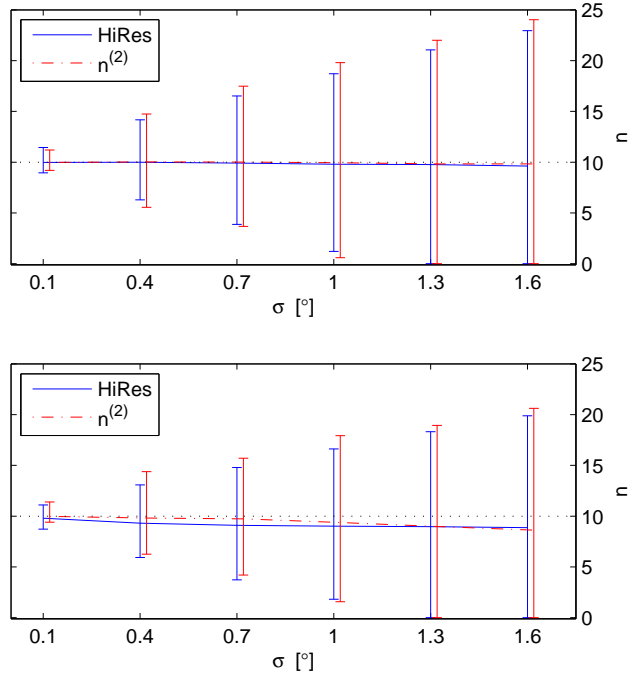}
 \end{center}
\caption{The number of correlations vs. the detector resolution. Error bars (slighly separated horizontally for readability) contain 90\% of the cases. \emph{Top:} Random sources; \emph{Bottom:} actual BLLac positions $-30^\circ \leq \delta \leq 90^\circ $.}
 \label{rand_bllacs}
 \end{figure}

We begin by testing the ability to reproduce the number of true correlations in the simplest case of dilute, random sources. A first simulation is done using half the sky, uniform detector exposure $R(\textbf{x})$, 156 randomly distributed sources and 271 cosmic rays (the numbers relevant to the BLLac studies of \cite{TT:BLLacs04} and \cite{HR:BLLacs}). Ten of the cosmic rays are Gaussianly aligned (a cosmic ray paired to a source, with angular separation according to the probability density $Q$, taken to be a $2d$ Gaussian of width $\sigma$), for various event resolutions. A second simulation uses the actual BLLacs source positions; these are more clustered than the random case. As shown in figure \ref{rand_bllacs}, both methods reproduce well on average the correct number of correlations.  The error bars shown include 90\% of the 10k realizations, and are similar for both methods. As $\sigma$ increases, the variance in the number of found correlations increases rapidly.

\subsection{Dispersion in results}

\begin{figure}
 \begin{center}
\includegraphics[scale=.75]{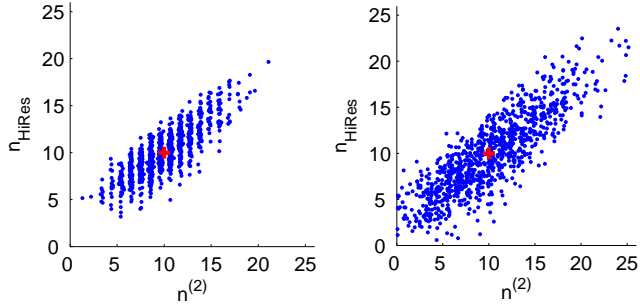}
 \end{center}
\caption{Comparison of the extracted number of correlations for specific realizations using the two methods. 1000 realizations of ten cosmic rays aligned to sources. \emph{Left:} For cosmic rays with event resolution $\sigma=0.4^\circ$. The banding is due to using the integer implementation of the generalized method. \emph{Right:} $\sigma=0.8^\circ$.}
\label{disp} 
\end{figure}

To test the extent of agreement between the two methods in individual realizations,  we generate 156 source positions at random on one hemisphere, align 10 cosmic rays to 10 of the sources and distribute 161 cosmic rays at random, then determine the number of correlations identified, with both procedures.  Figure \ref{disp} shows the results of 1000 repetitions, for $\sigma=0.4^\circ$ and for $0.8^\circ$.  From this exercise we conclude that the two ML methods {\em do not} generally agree for individual realizations, in spite of the fact that both methods reproduce the correct value in the mean.

\subsection{High source and UHECR densities}

\begin{figure}
 \begin{center}
\includegraphics[scale=.75]{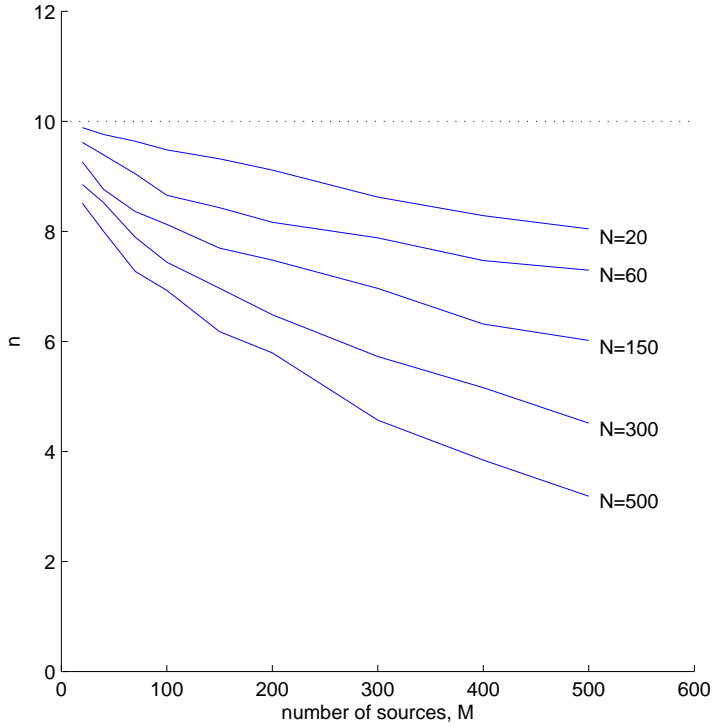}
 \end{center}
\caption{Correlations found by the HiRes method with 10 true correlations, $M$ sources and $N$ events in 100 square degrees. The method's prediction becomes increasingly inaccurate, as the dilute approximation becomes less valid.}
\label{dense} 
\end{figure}

When there are very large numbers of events and of potential sources, we expect the generalized ML method to perform worse than the HiRes method, beacause $n^{(2)}$ is obtained by taking the difference of two very large numbers, $n^{(0)}$ and $\bar{n}_{rand}$. Moreover, as the source density becomes very large it becomes impossible to reliably distinguish the ``contributing'' sources. The total number of correlations becomes the only interesting quantity to calculate. Thus, only the HiRes method should be used for the case of very high event densities. However, the HiRes method also deteriorates at high densities, as shown in figure \ref{dense}. 

\subsection{Sensitivity to experimental resolution}

\begin{figure}
 \begin{center}
 \includegraphics[scale=.75]{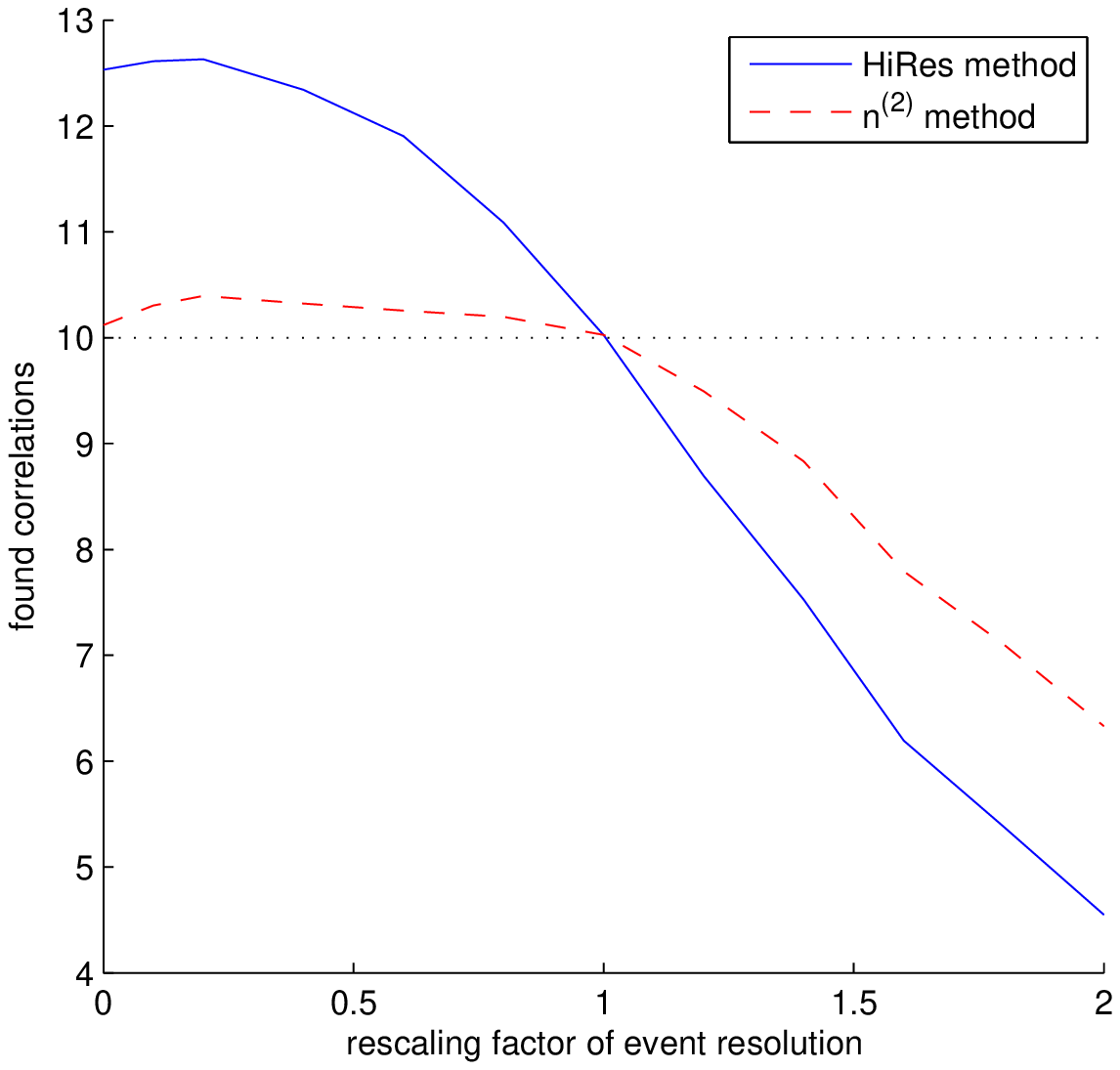}
 \end{center}
\caption{Average number of found correlations as a function of the factor by which $\sigma$ is rescaled by when ``Gaussianly aligning'' sources.}
 \label{skew}
 \end{figure}

If the resolution of cosmic ray events are consistently over- or underestimated in a given data set, the extracted correlations will be incorrect. To test the sensitivity of the two methods to this problem we repeat the first type of simulations, but rescale the event resolution when aligning a cosmic ray to a source. Figure \ref{skew} shows the average number of correlations found, as a function of the amount by which $\sigma$ is rescaled,  for the two methods. The new method is far less sensitive to incorrectly estimated resolution than is the HiRes method.

\subsection{Clustering of sources}

\begin{figure}
 \begin{center}
\includegraphics[scale=.75]{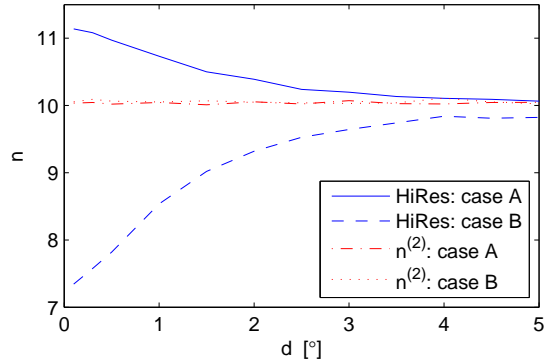}
 \end{center}
\caption{Sensitivity to clustering in source dataset, for UHECRs from (A) dense or (B) sparse regions.}
\label{clustering} 
\end{figure}

As seen in the lower panel of figure \ref{rand_bllacs}, spatial correlations within the data set of potential sources may skew the number of UHECR correlations found. In figure \ref{clustering} we show the results of a simulation with clustering of potential sources. The figure shows the mean for 10k realization of 271 cosmic ray events with $\sigma = 0.4^\circ$, and two different scenarios for the correlation with source clustering, for 156 candidate sources.   In both cases, candidate sources and CRs are distributed over one hemisphere; 100 sources are placed in ten randomly positioned clusters with ten sources distributed around each cluster center according to a $2d$ Gaussian of width $d$, and the remaining 56 candidate sources are placed at random in the hemisphere. In the first case, a randomly selected source in each cluster has one cosmic ray event Gaussianly aligned to it and the remaining 261 cosmic rays are placed at random. In the second case, ten CRs are Gaussianly aligned with ten of the randomly  placed source candidates \emph{not} in any cluster and the remaining CRs are placed at random. As figure \ref{clustering} demonstrates, the HiRes method significantly overestimates the true number of correlations if the sources are in clustered regions and underestimates it when the candidate sources show significant clustering but the UHECRs do not come from the clustered regions. By contrast the new method performs well in this test.

\section{Application to BLLacs and x-ray clusters}\label{applications}

In this section we apply both the HiRes method and the generalized method on two previously investigated data sets of suggested UHECR sources. 

\subsection{X-ray clusters}

The possibility that X-ray clusters are sources for UHECRs was investigated by \cite{pf05} using a binned analysis; a correlation was found between AGASA events and x-ray clusters taken from the NORAS catalog \citep{noras:2000}. Here we study the correlation between X-ray clusters and two samples of UHECRs, both separately and combined. We use the 375 X-ray clusters \citep{noras:2000} in the region $0^\circ<dec<80^\circ$, and $|b|>20^\circ$. For cosmic rays, we use the 143 HiRes events with $E>10^{19}$ eV, and 36 AGASA events with $E>4\times10^{19}$ eV, in the same  angular region. Note that in \cite{pf05} the authors restricted the correlation search to x-ray sources within the estimated GZK maximum distance for each UHECR event.   Imposing the GZK restriction reduces the number of found correlations in simulated random samples, but \cite{pf05} found that it does not significantly decrease the number of  correlations between UHECRs and x-ray clusters so that it increases the significance of the correlation.  In this paper we have not taken the GZK distance restriction into account, as our focus is on comparing the two ML methods.

When applying the Maximum Likelihood methods to x-ray clusters, we need to take the angular size of the clusters into account, as they can be of the same order as the cosmic ray event resolution. Therefore we introduce an effective event resolution to be used in the calculation of $Q$, $\sigma_{eff}=\sqrt{\sigma^2+r^2}$, where $r$ is the angular radius of the cluster. When combining the data sets we calculate the detector exposure by $R(\textbf{x})=\sum_iN_iR_i(\textbf{x})/\sum_iN_i$, with $i$ labelling the different data sets. For the generalized method we use the integer implementation.

The results of the various analyses are shown in table \ref{tab:n-xray}. We see that the two Maximum Likelihood methods are in fair agreement in all cases about the significance of the correlation, measured by $\mathcal{F}$.  However, the number of correlations inferred differs strikingly between the two methods for the combined HiRes-AGASA dataset. Perhaps surprisingly, the HiRes method finds more correlations in the AGASA sample alone than in the combined dataset. This is a hazard of using the HiRes method that we have probed by simulations. Consider the case of combining two sets of events, one with poor event resolutions with a number of events aligned to the set of sources, and the other set with good event resolutions but all events uncorrelated to the set of sources. In such a case the HiRes method significantly underestimates the number of correlated events. Conversely, if the aligned events all are taken from the set with good resolutions, the HiRes method overestimates the correct number of correlations. This explains the discrepancy between the methods for the combined data in table \ref{tab:n-xray}, since the AGASA events have poor event resolutions relative to the HiRes events. In effect, the HiRes method imposes a sort of internal consistency on the distribution of correlations, which could be inappropriate, if one data set had systematically incorrect $\sigma$ values for instance.

\begin{table}
\caption{\label{tab:n-xray}Correlations between X-ray clusters and UHECRs. The integer implementation has been used for the generalized method. For comparison, using the non-integer implementation we find $n^{(2)}=3.3$ and $n^{(2)}=9.6$ for the HiRes and AGASA event sets, respectively.}
\begin{indented}
\item[]\begin{tabular}{@{}lrrrlll}
\br 
  & \centre{3}{Generalized ML Method} &  & \centre{2}{HiRes Method} \\
UHECR & N & $n^{(0)}$ & $n^{(2)}$ & $\mathcal{F}$ & $n_{HR}$  & $\mathcal{F}_{HR}$ \\
\mr
 HiRes     &   143   &   22  &   5.2   &   0.3   &   1.5    &   0.3   \\
 AGASA     &    36   &   24  &   9.9   &   $6 \, 10^{-3}$   &   9.2    &   $9 \,10^{-3}$   \\
 Both   &   179   &   37  &  12.7   &   0.06   &   4.6    &   0.1 \\
\br
\end{tabular}
\end{indented}
\end{table}

\subsection{BLLacs}

The binned analysis performed in \cite{TT:BLLacs04} on the sample of 156 BLLacs with optical magnitude $V<18$ from the Veron 10th Catalog \citep{veron10} and the 271 HiRes events with $E>10^{19}$ eV showed a correlation at the $10^{-3}$ level. Applying the HiRes method on this sample we find $n_s=8.5$ correlations, and the fraction $\mathcal{F}$ of Monte Carlo runs with greater likelihood than the real data to be $2\times 10^{-4}$\footnote{This $n_s$ differs from the value ($n_s=8.0$) quoted in \cite{HR:BLLacs} probably due to the HiRes exposure map provided to us by HiRes (S. Westerhoff, private communication) being slightly different than the one used in the published HiRes analysis. For purposes of comparison of the two methods we do not require a perfect exposure map.}. Analyzing the same UHECR-BLLac data with the generalized ML method we find the number of correlations is $n^{(2)}=9.2$ with the integer implementation and 9.3 with the non-integer implementation, and $\mathcal{F}=6\times 10^{-5}$ for both. In fact from figure \ref{rand_bllacs}, we see that the HiRes method underestimates $n_s$ by 0.7 on average under these conditions, so on average this is the expected discrepancy. However as seen from figure \ref{disp}, the difference between the estimated number of correlations is also consistent with the variance from one realization to the next. With $1\sigma$ error bars, our best estimate is $9.2\pm2.5$ correlations.

Now including AGASA events too, in Table \ref{tab:n-bl} we list the results for the generalized method (integer implementation) and the HiRes method when applied to HiRes and AGASA data, both separately and combined. We note that while the calculated number of correlations differ between the methods, the difference is within the expected dispersion.

One of the advertised benefits of the generalized ML method is that it provides a way to rank sources according to the likelihood of them being correlated with cosmic ray events. To exemplify this we list in Table \ref{tab:bl} the individual BLLacs and correlated HiRes cosmic rays ranked according to the non-integer number of correlations found per source (cf. Table 1 in \cite{TT:BLLacs04}). The list includes all cosmic rays with positive $\ln \mathcal{R}$, which brings the number of cosmic rays listed to 16. Note that the sixth listed cosmic ray is not deemed a correlation by the generalized method, but is kept in the list to clarify why the corresponding source is found to have $n_j>1$ by the non-integer implementation of the method. Finally, we note that it would be desirable to impose restrictions on the sample of BLLacs to include only those within an estimated GZK maximum distance for each cosmic ray event, but unfortunately the redshift is known for only a fraction of the BLLacs.

\begin{table}
\caption{\label{tab:n-bl}Correlations between BLLacs and UHECRs. The integer implementation has been used for the generalized method.}
\begin{indented}
\item[]\begin{tabular}{@{}l r r r l l l}
\br 
  & \centre{3}{Generalized ML Method} &  & \centre{2}{HiRes Method} \\
UHECR & N & $n^{(0)}$ & $n^{(2)}$ & $\mathcal{F}$ & $n_{HR}$  & $\mathcal{F}_{HR}$ \\
\mr
HiRes     &   271   &   15  &   9.2   &   $6 \, 10^{-5}$  &   8.5    &   $2 \, 10^{-4}$\\
AGASA     &    57   &   13  &   6.4   &   0.04            &   2.6    &   0.1\\
Both   &   328   &   21  &  12.0   &  $3 \, 10^{-5}$   &  10.8    &   $1 \, 10^{-4}$\\
\br
\end{tabular}
\end{indented}
\end{table}

\fulltable{\label{tab:bl}HiRes UHECR cross correlations with BL Lacs using the noninteger implementation of the generalized Maximum Likelihood method.}
\br
\centre{3}{BLLacs} &  &  & \centre{4}{UHECRs} \\
  \bs
\crule{4}  &  & \crule{4} \\
\ms
Name & z & V & $n_j$ & $\Delta\theta$ ($^\circ$)  & $\ln \mathcal{R}$ & E (EeV)  & RA ($^\circ$) &  dec ($^\circ$) \\
\mr
SBS 1508+561       &  ...  &  17.3  &  1.905   &  0.83  &  3.18  &   21.7  &  226.56  &  56.61  \\
                   &           &        &          &  0.89  &  2.90  &   11.0  &  228.86  &  56.36  \\ 
MS 10507+4946      &   0.140   &  16.9  &  1.858   &  0.60  &  4.20  &   64.7  &  162.61  &  49.22  \\
                   &           &        &          &  1.04  &  2.08  &   24.1  &  165.00  &  49.29  \\
GB 0751+485        &  ...  &  17.1  &  1.268   &  0.32  &  4.33  &   12.0  &  118.77  &  48.08  \\
                   &           &        &          &  1.37  &  0.21  &   15.7  &  120.10  &  47.40  \\
1ES 1959+650       &   0.047   &  12.8  &  0.991   &  0.12  &  4.61  &   16.4  &  300.28  &  65.13  \\
FIRST J11176+2548  &   0.360   &  17.92 &  0.990   &  0.12  &  4.60  &   30.2  &  169.31  &  25.89  \\
Ton 1015           &   0.354   &  16.50 &  0.986   &  0.37  &  4.24  &   13.3  &  137.22  &  33.54  \\
RXS J01108-1254    &   0.234   &  17.9  &  0.986   &  0.37  &  4.24  &   10.5  &  \017.79  & \-12.56  \\
RXS J03143+0620    &  ...  &  17.9  &  0.980   &  0.50  &  3.89  &   27.1  &   \048.53  &   \05.84  \\
RXS J13598+5911    &  ...  &  17.9  &  0.970   &  0.62  &  3.47  &   24.4  &  210.02  &  59.80  \\
RGB J1652+403      &  ...  &  17.3  &  0.958   &  0.70  &  3.15  &   16.5  &  253.62  &  39.76  \\
RXS J08163+5739    &  ...  &  17.34 &  0.950   &  0.74  &  2.99  &   20.7  &  123.78  &  56.93  \\
OT 465             &  ...  &  17.5  &  0.849   &  0.95  &  1.98  &   10.5  &  265.37  &  46.72  \\
1ES 2326+174       &   0.213   &  16.8  &  0.527   &  1.12  &  1.08  &   38.1  &  352.40  &  18.84 \\
\br
\endfulltable

\section{Summary and Conclusions}\label{summary}

We have introduced a generalization of the HiRes Maximum Likelihood method, which allows the most likely sources of individual events to be identified and ranked.  Using simulations we have tested the two Maximum Likelihood methods and find that they complement each other well: the HiRes method allows a fast way to estimate the number of true correlated events, while the new method gives the quality of correlation between individual sources and cosmic rays rather than just the total number of correlated events.  Furthermore, the new method is less sensitive to the validity of the estimated angular resolution, and has less systematic bias when candidate sources are clustered (as astrophysical sources such as BLLacs are). Of course, for any given data set, mere statistical fluctuations can result in conclusions that would not be borne out with a larger data set. We conclude that both methods should be used: if they disagree markedly on the total number of correlations the data may have some statistical anomaly and be difficult to interpret. 

Although we have tested both Maximum Likelihood methods' ability to reproduce the true number of correlations under many different conditions, it is impossible to test for every conceivable statistical property of future data sets to which these methods could be applied. Thus we recommend, when applying these methods on new data samples, that the methods first be tested by simulations using the candidate source positions but randomly generated cosmic ray directions with some fraction of them Gaussianly aligned to some randomly chosen sources.  By repeating this procedure, the variance and possible systematic bias in the extracted number of correlations can be determined for that particular detector exposure, source and event statistics, resolution, etc.

We applied the ML methods to find correlations between x-ray clusters and AGASA events with $E>4\times 10^{19}$ eV, as done in \cite{pf05} but without imposing the GZK horizon.  We find $\approx 10$ excess correlations, with a chance probability of $6\times 10^{-3}$.  The same analysis for HiRes data with $E>10^{19}$ indicates $3-5$ excess correlations, which is expected by chance about 30\% of the time.   Imposing the restriction that the correlated x-ray cluster not be farther than the GZK distance cutoff was found in \cite{pf05} to reduce the number of chance correlations without significantly decreasing the number of true correlations.  Since in this work we did not impose the GZK restriction, we cannot compare the present analysis with that of \cite{pf05}.

Finally, we applied the generalized Maximum Likelihood method to check previously claimed correlations between UHECRs and BLLacs.  We corroborate that there is a significant correlation between BLLacs of the Veron 10th catalog and HiRes cosmic rays with $E>10^{19}$ eV \citep{HR:BLLacs}.  The generalized ML method yields a slightly greater number ($\approx 9$ instead of $\approx 8$) excess correlations and a lower chance probability ($6 \times 10^{-5}$ instead of $2 \times 10^{-4}$).  Simulations show that the  HiRes ML method systematically underestimates the number of correlations by 0.7 when the source candidates are clustered like the BLLacs, suggesting that the present result is the more reliable. The striking difference between the apparently significant correlation observed in the Northern sky (HiRes) and Auger's null result \citep{Harari:2007up} is a mystery. It may be due to greater average magnetic deflection in the Southern sky, or greater incompleteness in the Southern BLLac catalog, or a statistical fluctuation or a combination of the above.

\ack
We thank S. Westerhoff, C. Finley, E. Pierpaoli and I. Zaw for information and discussions.  This research has been supported in part by NSF-PHY-0401232.

\bibliographystyle{jphysicsB}
\bibliography{ML041608jcap}

\end{document}